# Mathematical Modelling and Comparative Study of Elliptical and Circular Capacitive Pressure Microsensor

**Rishabh Bhooshan Mishra, S. Santosh Kumar***
Process Technologies Group, Smart Sensor Area, CSIR – Central Electronics Engineering Research Institute (CEERI), Pilani, Rajasthan, India – 333031

*Email: santoshkumar.ceeri@gmail.com

**Abstract.** A lot of work on clamped circular, square and rectangular shaped capacitive pressure sensor has been carried out. However, to the best of our knowledge, no elaborate work has been performed on mathematical formulation and simulation of clamped elliptical shaped capacitive pressure sensor in literature. This paper describes the mathematical modelling and simulation of normal mode clamped elliptical shaped capacitive pressure sensor. The study of various performance parameters like maximum diaphragm deflection, capacitance variation, mechanical sensitivity, capacitive sensitivity and non-linearity is also carried out for operating pressure range of 0kPa – 18 kPa. For circular capacitive pressure sensor, the operating pressure range is modified according to physical dimensions i.e. thickness and radius of diaphragm, separation gap between plates and maximum deflection. For same overlapping area between plates ($10000\pi$ mm$^2$), the comparison of elliptical shapes of different eccentricities with circular shape diaphragm is carried out. The thickness of diaphragm is taken as 2 μm and separation gap is 1 μm in all the designs which are used in this work. In this comparative study, it is observed that elliptical shaped capacitive pressure sensors have better linearity than circular diaphragm pressure sensor.

## 1. Introduction

Micro electromechanical system (MEMS) is popular and developed technology, adopted from IC-processing, for fabricating the micro-sensors for various process variable measurements like pressure, temperature, distance, acceleration, fluid flow, rotation and angular velocity etc. In general, MEMS devices consist of moveable micromechanical components like bridges, beams, gears, diaphragms, channels, micro-pumps, valves, cantilevers which are integrated or interfaced with electronic components or devices. The Silicon based devices/structures are micro-fabricated by a particular process flow sequence of photolithography, etching and deposition to fabricate desired components/structure. The integration of microelectronic and mechanical components in a single chip makes the MEMS devices more versatile than conventional devices. To apply conventional solid mechanics theories and practices to the MEMS based systems are interesting and challenging task. For measuring the absolute, gauge or differential pressure using piezo-resistive and capacitive pressure sensors, the MEMS based sensors utilize movement in diaphragm due to pressure application [1-4].

In piezo-resistive pressure sensor, four piezo-resistors are mounted on the diaphragm and connected in whetstone bridge fashion. After pressure application, there is deflection in the diaphragm on which piezo-resistors are mounted. The change in resistance of implanted piezo-resistors unbalances the Wheatstone bridge which provides the output voltage according to the applied pressure [5-6]. However, due to several advantages of capacitive pressure sensor over piezo-resistive pressure sensor like long term stability, less power consumption and temperature drift etc., the capacitive pressure sensors are used in several applications like biomedical, aerospace, consumer electronics and automobile. Achieving linearity characteristic and sealing vacuum cavity is a major tough task in fabrication of the capacitive pressure sensor. Since there is no need to mount any piezo-resistor in the capacitive sensor diaphragm, so scaling the dimension of the diaphragm is bit easier. However, nonlinearity between applied pressure vs. capacitance change, large impedance of output signal of sensor, small change in capacitance and parasitic capacitance between output of device and capacitance measurement circuitry are some of the major disadvantage of capacitive pressure sensor. Some issues must be addressed by circuitry which is used to measure the change in capacitance. As sealing the vacuum cavity is complex task, so proper care must be been taken at the time of fabrication and packaging the absolute capacitive pressure sensor [7-9].

Bio-MEMS is, one of revolutionized area of MEMS application, utilizing VLSI or micro-fabrication technology for implantable/physiological devices because of several advantages. In living beings, several implantable devices are based on MEMS technology like Biosensors for in-vivo sensing, immune-isolation devices, drug delivery systems (micro-reservoir, micro-particle) and some inject able devices like micro-needle, micro-module. Subsequently, the data obtained from the Bio-MEMS devices is transmitted using wireless communication technology for study of various body-parts or continuous monitoring purpose. In medical sector, if any part of living-being dysfunctions then several steps can be taken to save the life [10-11].

The general capacitance measurement circuitry for measuring capacitive pressure sensor is astable mutivibrator which converts the capacitance variation (due to applied pressure) into frequency. The astable mutivibrator contains a LF351 Op-amp and three external resistances. And MS3110 is the commercially available Integrated Circuitry for capacitance variation measurement which converts capacitance variation into voltage [12]. The algometric capacitance to digital converters can also be utilized for measuring the base capacitance and variation in capacitance after pressure application [13].

## 2. Paper Structure

The sufficient amount of work like design, modelling, simulation and fabrication of square and circular shaped capacitive pressure sensors have been performed earlier. However, as per best of our literature review, no mathematical modelling and numerical simulation of clamped elliptical shaped capacitive pressure sensor is capacitive pressure sensor. The simulation and fabrication of L-shaped clamped elliptical capacitive pressure sensor, with signal conditioning circuitry, have been reported [14]. The finite element analysis (FEA) of elliptical shaped capacitive pressure sensor, with temperature variation, has been reported [15]. However the mathematical analysis and study of various performance characteristics with numerical analysis is unavailable.

In the presented work, elliptical shaped diaphragms of different eccentricities are chosen and compared with the circular shaped diaphragm. The overlapping area between plates is kept constant for all sensor designs. In all, six different designs consisting of one circular shaped and five elliptical shaped capacitive pressure sensors of 2 μm thicknesses and 1 μm separation gap are chosen for numerical simulation. Overlapping area of different designs are same i.e. 10000π μm$^2$. The specifications of six different designs are listed in Table 1. Where, $a$ and $b$ are semi-major and semi-minor axes of ellipse and L is radius of circle.

**Table 1.** Dimensions of Specified Sensors.

| S. N. | Diaphragm Shape | Diaphragm dimensions (μm) |
|---|---|---|
| 1. | Circular | L = 100 |
| 2. | Elliptical | a = 125, b = 80 |
| 3. | Elliptical | a = 200, b = 50 |
| 4. | Elliptical | a = 250, b = 40 |
| 5. | Elliptical | a = 500, b = 20 |
| 6. | Elliptical | a = 1000, b = 10 |

## 3. Deflection Theory and Kirchhoff's assumption

The deflection theory is based on small and large deflection in diaphragm/plate. To apply small deflection theory, the diaphragm deflection must be less than 1/5[th] of diaphragm thickness. In case of large deflection theory, the diaphragm deflection must be five times of diaphragm thickness or more. The small deflection theory of plate is utilized for normal mode capacitive pressure sensor and large deflection theory can be utilized for touch or double touch mode capacitive pressure sensor [4, 7-8].

The Small deflection theory is based on certain Kirchhoff's assumptions, which are:

- The first assumption is based on material property of diaphragm. The diaphragm must be made of elastic, homogeneous and isotropic material.
- The second assumption deals with the dimension or geometry of diaphragm. The diaphragm must be thin as well as flat. The dimension of diaphragm must be larger than ten times than diaphragm thickness.
- The third assumption states about the maximum diaphragm deflection, which must not be larger than 1/5[th] of diaphragm thickness.
- The fourth assumption is known as 'hypothesis of straight normal' which states that the middle plane of diaphragm does not have normal stress and should not be strained. The strain lines should remain straight and perpendicular to the middle plane.
- The middle plane of plate remains unstrained after bending.

## 4. Analytical Modeling

*4.1 Deflection in Clamped Elliptical Diaphragm*

The boundary equation of classical elliptical shaped plate, shown in Figure 1, can be given by:

$$\left(\frac{x}{a}\right)^2 + \left(\frac{y}{b}\right)^2 = 1 \tag{1}$$

The eccentricity, *e*, of the elliptical boundary can be given by (if b < a):

$$e = \sqrt{1 - \left(\frac{b}{a}\right)^2} \tag{2}$$

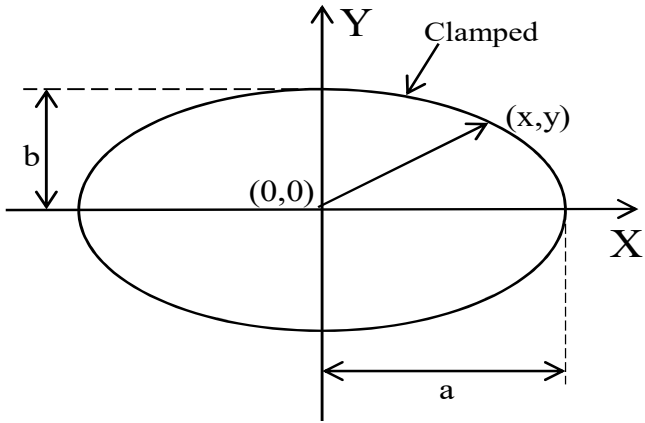

**Figure 1.** The basic capacitive pressure sensor consists two parallel plates which are separated by a media. In which one plate will be fixed and another is moveable after pressure application. This shown figure is top view of capacitive pressure sensor which consist elliptical shaped diaphragm.

If uniform pressure P is applied, in normal direction, to clamped elliptical shaped Kirchhoff's plate of thickness, Young's modulus of elasticity and Poisson's Ratio t, E and ν, respectively, then plate equation is expressed by a fourth order partial differential equation which can be given by [16-17]:

$$\Delta\Delta\,[w(x,y)] = \frac{P}{D} \tag{3}$$

here, flexural rigidity of plate is $D = Et^2/(12 - v^2)$, $w(x, y)$ is deflection at any point (x, y) on plate and $\Delta$ is Laplace operator or Laplacian. The deflection in plate is perpendicular to 2D – plane. In 2D co-ordinate $\Delta$ can be given by:

$$\Delta \equiv \nabla^2 = \frac{\partial^2}{\partial x^2} + \frac{\partial^2}{\partial y^2} \tag{4}$$

The applied boundary conditions to solve the eq. (3) are:

$$w = 0 \tag{5}$$

and,

$$\frac{\partial w}{\partial n} = 0 \tag{6}$$

here, the normal to elliptical plate edge is $n$.
Using both the boundary conditions, deflection in clamped elliptical plate can be given by [16-17]:

$$w_e(x,y) = w_{e,0}\left[1 - \left(\frac{x}{a}\right)^2 - \left(\frac{y}{b}\right)^2\right]^2 \tag{7}$$

The maximum deflection due to pressure application is found at the centre of plate and this can be obtained after putting x = 0 and y = 0 in eq. (7). Therefore, maximum deflection in the clamped elliptical shaped diaphragm for $a > b$ can be given by [16-17]:

$$w_{e,0} = \frac{P}{8D\left(\frac{3}{a^4} + \frac{3}{b^4} + \frac{2}{a^2 b^2}\right)} \tag{8}$$

In case of zero eccentricity i.e. $a = b = L$, the given formula of deflection in elliptical diaphragm can be modified for clamped circular shaped diaphragm which can be given by [4,7-9,12, 16-17]:

$$w_c = w_{c,0}\left[1 - \left(\frac{r}{L}\right)^2\right]^2 \tag{9}$$

here, $w_{c,0}$ is the maximum deflection in circular shaped plate which can be given by:

$$w_{c,0} = \frac{PL^4}{64D} \tag{10}$$

From eq. (8) and (10), it is clear that the maximum deflection, in clamped elliptical and circular both, is the function of applied pressure, dimensions of plate, material properties and plate thickness.

*4.2 Base Capacitance of Sensor*
If permittivity of medium and separation gap between plates is ε and d, respectively, then base capacitance of sensor can be given by []7-9, 12:

$$C_b = \frac{\varepsilon \times \text{overlapping area between plates}}{\text{separation gap}} \tag{11}$$

The overlapping area of elliptical shaped plate can be given by:

$$A_e = \pi ab \tag{12}$$

And overlapping area of circular shaped plate can be given by:

$$A_c = \pi L^2 \tag{13}$$

*4.3   Capacitance variation in sensor*

The capacitance in terms of applied pressure P on elliptical plate can be given by double integral [4, 7]:

$$C_{w,e} = \iint_G \frac{\varepsilon\, dx\, dy}{d - w(x,y)} \tag{14}$$

here, region $G$ is: $\left(\frac{x}{a}\right)^2 + \left(\frac{y}{b}\right)^2 = 1$.

Since the term which depends on applied pressure i.e. $d - w(x, y)$, is in denominator of capacitance variation, so the capacitance increases as diaphragm deflection increases.

From Eq. (7), (8) and (14):

$$C_{w,e} = \iint_G \frac{\varepsilon\, dx\, dy}{d - w_{e,0}\left[1 - \left(\frac{x}{a}\right)^2 - \left(\frac{y}{b}\right)^2\right]^2} \tag{15}$$

After applying transformation: $x = au$ and $y = bv$ [18]:

$$C_{w,e} = \varepsilon ab \iint_H \frac{du\, dv}{d - w_{e,0}[1 - u^2 - v^2]^2} \tag{16}$$

here, region H is: $u^2 + v^2 = 1$.

After applying transformation theory of cylindrical coordinate, $u = q\cos\varphi$, $v = q\sin\varphi$ and $z = z$, we get:

$$C_{w,e} = \varepsilon ab \iint_I \frac{q\, d\varphi\, dq}{d - w_{e,0}[1 - q^2]^2} \tag{17}$$

here, region $I$ is: $q^2 = 1$.

After solving this double integral, we get:

$$C_{w,e} = \frac{\varepsilon A_e}{2\sqrt{dw_{e,0}}} \ln\left|\frac{\sqrt{d} + \sqrt{w_{e,0}}}{\sqrt{d} - \sqrt{w_{e,0}}}\right| \tag{18}$$

In case of zero eccentricity i.e. for the circular diaphragm (when $a = b = L$), the capacitance after pressure application can be given by [4, 7-9]:

$$C_{w,c} = \frac{\varepsilon A_c}{2\sqrt{dw_{c,0}}} \ln\left|\frac{\sqrt{d} + \sqrt{w_{c,0}}}{\sqrt{d} - \sqrt{w_{c,0}}}\right| \tag{19}$$

*4.4 Sensitivity of Sensor*

The mechanical sensitivity of capacitive pressure sensor can be obtained by differentiating deflection at the centre i.e. maximum deflection w.r.t. pressure which plays very important role in optimization of sensor designs, if the maximum deflection for few sensor designs is approximately same. Then mechanical sensitivity of elliptical shaped capacitive pressure sensor can be given by:

$$S_{e,mech} = \frac{\partial w_{e,0}}{\partial P} \tag{20}$$

After differentiating the Eq. (8), we get:

$$S_{e,mech} = \frac{1}{8D\left(\frac{3}{a^4} + \frac{3}{b^4} + \frac{2}{a^2 b^2}\right)} \tag{21}$$

The mechanical sensitivity of capacitive pressure sensor is function of dimension and flexural rigidity of plate.

The capacitive sensitivity of capacitive pressure sensor is obtained by differentiating capacitance change w.r.t. pressure range. Then capacitive sensitivity of elliptical shaped capacitive pressure sensor can be given by [8-9]:

$$S_{e,cap} = \frac{\partial C_{w,e}}{\partial P} \tag{22}$$

After performing differentiation, we get:

$$S_{e,cap} = \frac{S_{e,mech}}{2w_{e,0}}\left(\frac{\varepsilon A_e}{d - w_{e,0}} - C_{w,e}\right) \tag{23}$$

*4.5 Non-linearity in Sensor*

The non-linearity of a sensor is defined as the deviation of ideal curve with obtained output curve of sensor. This ideal curve is obtained using end point straight line or least point fit curve. For each point on ideal curve there will be particular deviation in the output curve. And the maximum deviation of output curve w.r.t. ideal curve (in percentage) is called non-linearity. The non-linearity, at any specific point, of elliptical shaped capacitive pressure sensor is given by [5,9]:

$$NL_i(\%) = \frac{C_{w,e}(P_i) - C_{w,e}(P_m) \times \frac{P_i}{P_m}}{C_{w,e}(P_m)} \times 100 \tag{24}$$

here, $P_i$ is the pressure at ideal curve, $P_m$ is the maximum applied pressure, $C_{w,e}(P_i)$ is the output capacitance at pressure $P_i$ on ideal curve and $C_{w,e}(P_m)$ is output capacitance at pressure $P_m$ on ideal curve.

## 5. Simulation Results and Discussion

The maximum deflection in circular and elliptical shaped diaphragm at different pressures is shown in Figure 2. The diaphragm deflection is kept less than 1/5th of diaphragm thickness (i.e. 0.4 μm) and 1/4th of separation gap (i.e. 0.25 μm) to follow small deflection theory of plates and avoid pull-in phenomena, respectively. The operating pressure range is 0kPa to 18 kPa. Silicon is used as the diaphragm material; it has a Young's Modulus of elasticity of 169.8 GPa and Poisson ratio of 0.066. The deflection is found to be more in the case of circular diaphragm than in the elliptical shaped diaphragms. All the designs have linear defection curve in applied pressure range which follows the Hook's Law. However, as the value of semi-major axis increases and semi-minor axis of elliptical diaphragm decreases, keeping constant the overlapping area between plates, the deflection in plate decreases.

Figure 3 represents 2-D plot of deflection in elliptical shaped clamped diaphragm of designed sensor. In this 2-D plot, the deflection is maximum at centre of diaphragm and then decreases as we move towards the clamped edges.

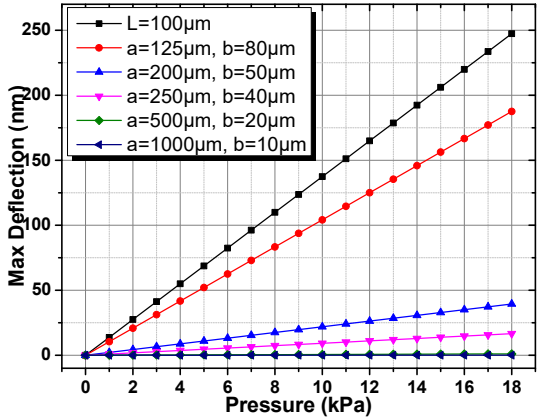
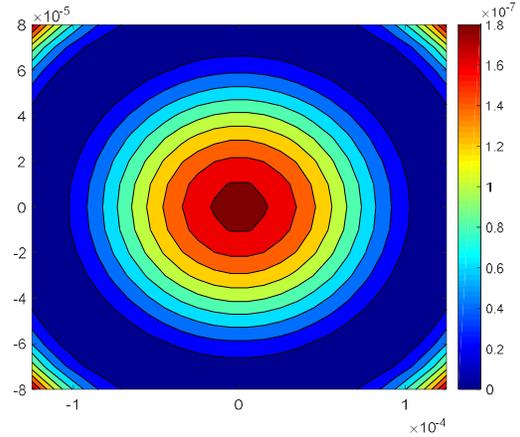

**Figure 2.** Deflection in various diaphragms of thickness 2 μm.

**Figure 3.** Deflection plot in elliptical diaphragm of a = 125 μm and b = 80 μm of 2 μm thickness at 18 kPa Pressure.

The overlapping area of all the proposed designs are kept same, so base capacitance (capacitance without bending of diaphragm due to applied pressure) of all the designs are 0.27816 pF. The capacitance variation in different sensor designs, w.r.t. applied pressure range of 0 kPa – 18 kPa, is given in Figure 4. The capacitance change in circular capacitive pressure sensor is found maximum. The capacitance change decreases by decreasing semi-minor axis and increasing semi-major axis, keeping the overlapping area between plates constant.

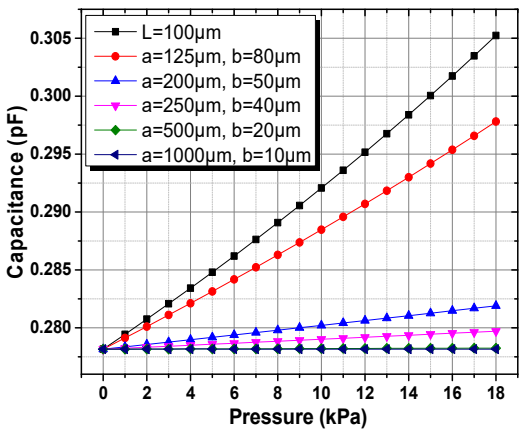
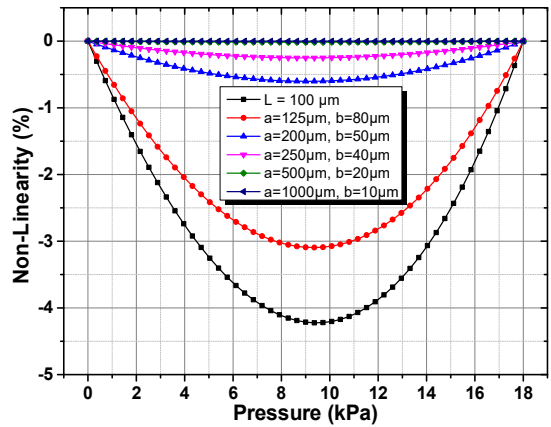

**Figure 4.** Capacitance variation in sensors of specified dimensions.

**Figure 5.** Non-linearity in sensors of specified dimensions.

The mechanical and capacitive sensitivity of different diaphragm dimensions are shown in Table 2. Mechanical sensitivity can be important factor, if the deflection due applied pressure is different in case of different design. The mechanical sensitivity of circular shaped capacitive pressure sensor is largest among all designs and decreases with increasing the value of semi-major axis and decreasing the value of semi-minor axis in case of elliptical capacitive pressure sensor.

The capacitive sensitivity of circular plate is highest amongst all proposed designs. The capacitive sensitivity decreases in case of elliptical capacitive pressure sensor as value of semi-major axis increases and semi-minor axis decreases, keeping the overlapping area between plates constant.

Figure 5 shows the non-linearity of different designs. Non-linearity is one of important parameter for pressure sensors. The problem of the non-linearity can be controlled by reducing large deflection of diaphragm centre. The non-linearity in circular shaped capacitive pressure sensor is maximum. In case of elliptical shaped capacitive pressure sensor, non-linearity reduces as the semi-major axis increases and correspondingly semi-minor axis decreases.

**Table 2.** Mechanical and Capacitive Sensitivity of Specified Sensors.

| S. N. | Diaphragm Dimension (μm) | Mech. Sensitivity (μm/kPa) | Cap. Sensitivity (fF/kPa) |
|---|---|---|---|
| 1. | L = 100 | $1.3743 \times 10^{-2}$ | 1.5044 |
| 2. | a = 125, b = 80 | $1.1042 \times 10^{-2}$ | 1.0916 |
| 3. | a = 200, b = 50 | $2.1906 \times 10^{-3}$ | $2.0778 \times 10^{-1}$ |
| 4. | a = 250, b = 40 | $9.2184 \times 10^{-3}$ | $8.611 \times 10^{-2}$ |
| 5. | a = 500, b = 20 | $5.8574 \times 10^{-5}$ | $5.4343 \times 10^{-3}$ |
| 6. | a = 1000, b = 10 | $3.6645 \times 10^{-6}$ | $3.3978 \times 10^{-5}$ |

## 6. Conclusion

In this work, the mathematical modelling and simulation for elliptical shaped capacitive pressure sensor is carried out and the various parameters for pressure sensors are compared with circular shape sensor, while the overlapping area between plates is kept same. To the best of our knowledge, there is no elaborate work in literature about elliptical shaped capacitive pressure sensor. The various performance parameters have been derived step by step using mathematical modelling and MATLAB® is utilized for validating the mathematical modelling. In performing simulations, the overlapping area between plates is kept constants for all specified dimensions of elliptical and circular shape.

The circular diaphragm has more deflection, higher change in capacitance, better mechanical and capacitive sensitivities in comparison to other elliptical diaphragms even though all of them have same overlapping area between plates. However, it also has the highest non-linearity of 4.22%. The elliptical diaphragm of dimension a = 125 μm and b = 80 μm has maximum deflection of 0.24737 μm at maximum applied pressure is 18 kPa, change in capacitance at full scale value of pressure is 21.49 fF, mechanical sensitivity is $1.10418 \times 10^{-2}$ μm/kPa, capacitive sensitivity is 1.09167 fF/kPa and maximum non-linearity is -3.089% which is 3/4th of circular shaped capacitive pressure sensor. Therefore, it is concluded that elliptical shaped diaphragms are better suited when non-linearity is an important consideration. Otherwise, the circular shaped diaphragm is more suitable in terms of all other parameters.

**Acknowledgement**

Authors acknowledge the Director, CSIR – CEERI, Pilani, for providing generous support. This presented analysis is also supported by Dr. Ankush Jain (Scientist, CSIR-CEERI). The authors wish to acknowledge all the scientific and technical staff of Process Technologies Group, Smart Sensor Area, CSIR-CEERI.